\documentclass{article}

\usepackage{arxiv}
\usepackage[utf8]{inputenc} 
\usepackage[T1]{fontenc}    
\usepackage{hyperref}       
\usepackage{url}            
\usepackage{booktabs}       
\usepackage{amsfonts}       
\usepackage{amsmath} 
\usepackage{nicefrac}       
\usepackage{microtype}      
\usepackage{lipsum}		
\usepackage{graphicx}
\usepackage{cite}
\usepackage{doi}
\usepackage{caption}
\usepackage{array}
\usepackage{multirow}
\usepackage{algorithm}
\PassOptionsToPackage{noend}{algpseudocode}
\usepackage{algpseudocode}

\algdef{SE}[DOWHILE]{Do}{doWhile}{\algorithmicdo}[1]{\algorithmicwhile\ #1}%

\title{When fire attacks: How does concrete stand up to heat?}

\author{ \href{https://orcid.org/0009-0004-4945-9821}{\includegraphics[scale=0.06]{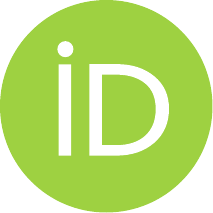}\hspace{1mm}Anshu Sharma}\\
	Department of Civil Engineering\\
	Indian Institute of Technology (BHU) Varanasi\\
	\texttt{anshusharma.rs.civ23@itbhu.ac.in} \\
	\And
	\href{https://orcid.org/0000-0002-4826-2232}{\includegraphics[scale=0.06]{orcid.pdf}\hspace{1mm}Basuraj Bhowmik*} \\
	Department of Civil Engineering\\
	Indian Institute of Technology (BHU), Varanasi\\
	\texttt{basuraj.civ@iitbhu.ac.in} \\
	*Corresponding author
}

\renewcommand{\shorttitle}{Performance based evaluation of concrete beams under fire loads}

\begin{document}
	\maketitle
	
	\begin{abstract}	
	Fire is a process that generates both light and heat, posing a significant threat to life and infrastructure. Buildings and structures are neither inherently susceptible to fire nor completely fire-resistant; their vulnerability largely depends on the specific causes of the fire, which can stem from natural events or human-induced hazards. High temperatures in structures can lead to severe health risks for those directly affected, discomfort due to smoke, and compromised safety if the structure fails to meet safety standards. Elevated temperatures can also cause significant structural damage, becoming the primary cause of casualties, economic losses, and material damage.
	
	This study aims to investigate the thermal and structural behavior of concrete beams when exposed to extreme fire conditions. It examines the effects of different temperatures on plain and reinforced concrete (PCC and RCC, respectively) using finite element method (FEM) simulations. Additionally, the study explores the performance of various concrete grades under severe conditions. The analysis reveals that higher-grade concrete exhibits greater displacement, crack width, stress, and strain but has lower thermal conductivity compared to lower-grade concrete. These elevated temperatures can induce severe stresses in the concrete, leading to expansion, spalling, and the potential failure of the structure. Reinforced concrete, on the other hand, shows lower stress concentrations and minimal strain up to 250°C. These findings contribute to the existing knowledge and support the development of improved fire safety regulations and performance-based design methodologies.
	\end{abstract}

\section{Data} 
\begin{figure}[h]
	\centering
	\includegraphics[width=0.8\textwidth]{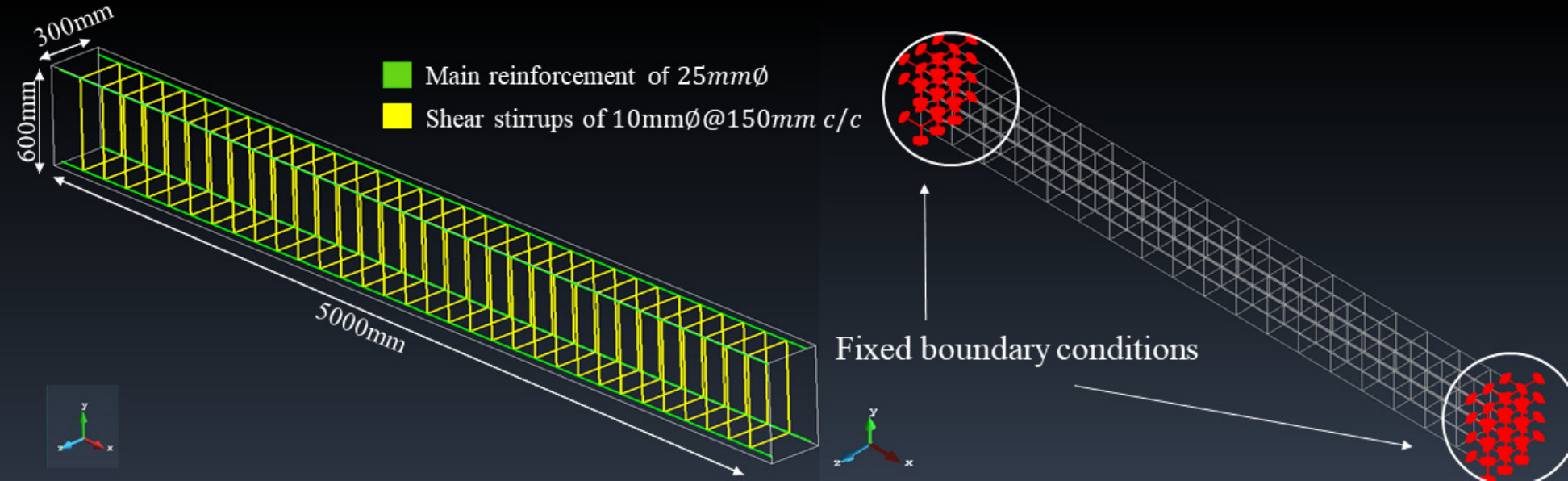}
	\caption{ATENA beam model with reinforcement details (left) and boundary conditions (right)}
	\label{figure1}
\end{figure}

The data presented in this study focuses on the performance of concrete under severe fire conditions, building on previous research in this field \cite{kodur2008numerical,alvarez2013twenty}. Four different concrete grades were examined at four distinct temperatures, utilizing a beam model designed and tested with ATENA (Advanced Tool for Engineering Nonlinear Analysis). ATENA is a FEM-based software that accurately simulates the real behavior of concrete and reinforced concrete structures, including cracking, crushing, and reinforcement yielding. Separate models for plain cement concrete (PCC) and reinforced cement concrete (RCC) beams were developed and tested under these conditions. Fig. \ref{figure1} illustrates the RCC beam model and the boundary conditions applied in the analysis. This model facilitated the assessment of concrete based on thermal conductivity, displacement, crack width, maximum stress, and maximum strain.

The detailed specifications of the data are outlined in Table \ref{tab:specstable}, which provides an overview of the subject area, data acquisition methods, analysis factors, and features. The analysis was conducted by raising the maximum temperature from its ambient $25^\circ$C to $100^\circ$C, $250^\circ$C, $500^\circ$C, and $750^\circ$C. For each temperature, analysis was performed on a fixed concrete beam of varying grades M25, M35, M45, and M50, subjected to self-weight. The thermal and structural results were recorded for different grades of plain and reinforced cement concrete (PCC and RCC respectively) to evaluate structural performances at various temperatures. The value of the data is highlighted in Table \ref{tab:valuetable}, emphasizing its significance as a benchmark for understanding the performance of concrete structures exposed to high temperatures.  

	\section{Executive summary}
\begin{table}[h!]
	\caption{Specifications table}
	\centering
	\begin{tabular}{ll}
		\toprule
		\multicolumn{2}{c}{}             \\
		\textbf{Subject area}     & \textit{Fire Analysis }   \\
		\textbf{More specific subject area  }   & \textit{Concrete behaviour under severe fire conditions}   \\
		\textbf{Type of data}     & \textit{Excel Datasheet }   \\
		\textbf{How data was acquired }   & \textit{By FEM analysis using ATENA/transport and ATENA/static}   \\
		\textbf{	Data format}    & \textit{Fire Analysis}    \\
		\textbf{Analysis factors }    &  \textit{The maximum temperature was raised from its ambient} $25^\circ$C \textit{to different temperatures viz.}, \\
		& $100^\circ$C, $250^\circ$C, $500^\circ$C, \textit{and} $750^\circ$C. \textit{For each temperature, analysis is performed on a fixed} \\ 
		& \textit{concrete beam of varying grades M25, M35, M45, and M50, subjected to self-weight.} \\
		\textbf{Analysis features}   & \textit{Thermal and structural results are recorded for different grades of plain and reinforced cement }\\ 
		& \textit{concrete (PCC and RCC respectively) to evaluate structural performances at various temperatures.  } \\
		\textbf{Data accessibility }   & \textit{Fire Analysis  }  \\
		\bottomrule
	\end{tabular}
	\label{tab:specstable}
\end{table}

\begin{table}[h]
	\caption{Value of the data}
	\centering
	\begin{tabular}{l}
		\toprule
		
		1. The data will provide a benchmark for understanding performance of concrete structures exposed to high temperatures. \\
		2. The data is expected to be an important resource for assessing output to overcome real challenges in fire prone areas. \\
		3. The data will serve as a key reference for future research in concrete fire analysis for difference compositions of grade \\ 
		and in further discoveries of alternate materials.\\
		
		\bottomrule
	\end{tabular}
	\label{tab:valuetable}
\end{table}

\section{Numerical fabrication, material, and methods}

\subsection{Beam Simulations}
The fire analysis procedure begins with preparing a beam model using ATENA-GiD, where geometrical parameters such as breadth (B), depth (D), and length (L) are assigned. Subsequently, global problem data, including the problem type, method, and solver, are defined in the ATENA program. The fire analysis is divided into thermal analysis and static analysis. The thermal results obtained are used for further calculations in the static analysis. For the thermal analysis, the ATENA/transport version is selected. Solution parameters, such as convergence criteria and iteration limits, are set and kept constant across all cases to ensure uniformity in the results. Material properties, boundary conditions, and loading conditions are then assigned. The fire load is applied using different fire growth curves that represent the pattern of fire growth (heat release rate) over time \cite{steen2021learning, kobes2010building, thevega2022fire}. In this case, the fire growth curve is defined by Eq. \ref{hceq}, representing a hydrocarbon curve \cite{eurocode2002actions}. 

\begin{figure}[h]
	\centering
	\includegraphics[width=0.8\textwidth]{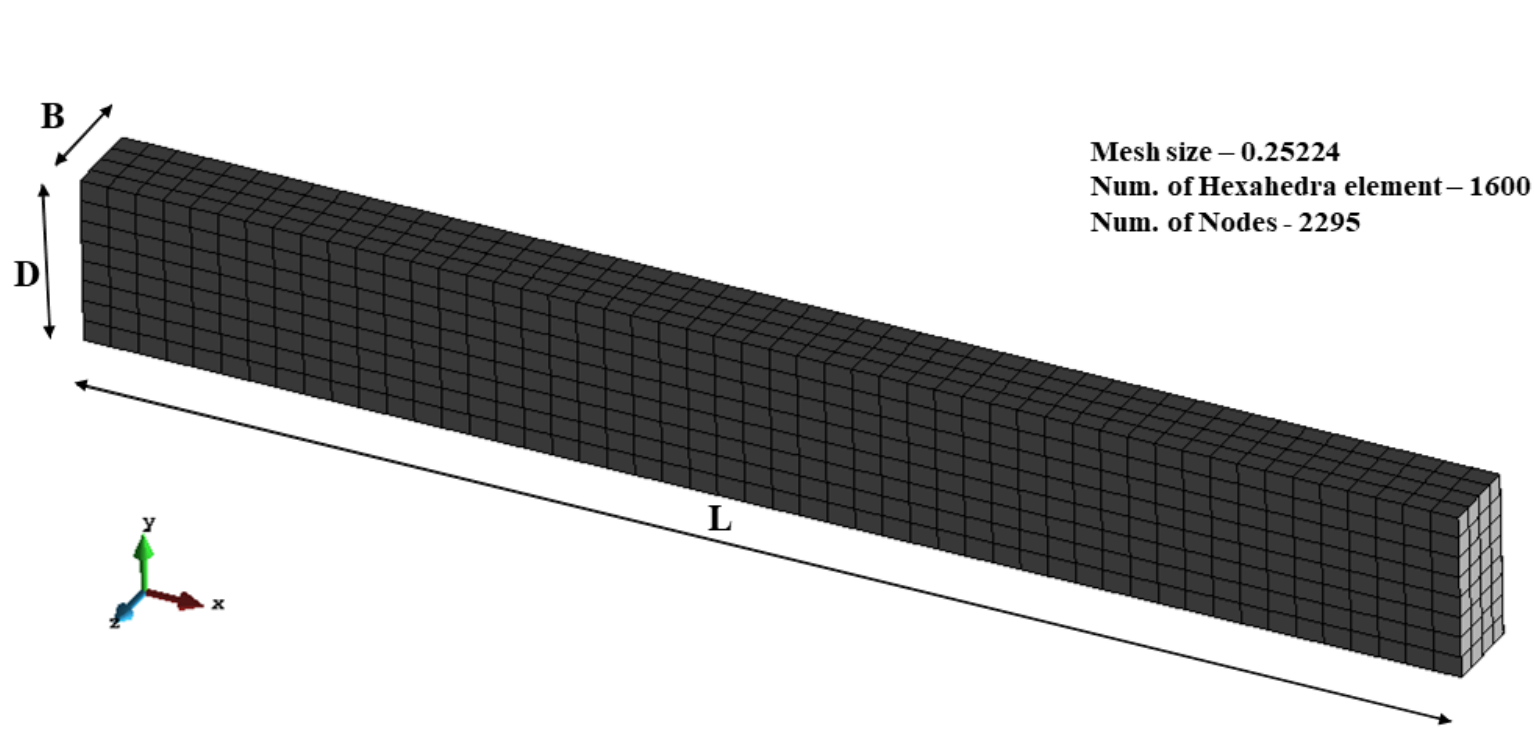}
	\caption{ATENA beam model with mesh representation}
	\label{figure2}
\end{figure}

\begin{equation}
	\label{hceq}
T = {T_0} + 1080\left( {1 - 0.325{e^{ - 0.167t}} - 0.675{e^{ - 2.5t}}} \right)
\end{equation}
The interval solution parameter is then defined for gradual application of loading because it is a generous practice to apply non-linear loads in sufficient intervals. After assigning all the parameters, mesh is generated as structured mesh using hexahedral elements as shown in Fig. \ref{figure2}. 
 
The convergence criteria are then checked against the convergence limit \cite{kodur2010high,mroz2016material,malik2021thermal}. If they are met, the thermal analysis yields $\psi$-values. If not, the iteration limits are increased or the mesh size is revised, and the process is reanalyzed. Upon completion of the thermal analysis, the thermal results are imported for static analysis using the ATENA/static version. The static analysis follows similar steps to the thermal analysis, including the definition of solution parameters, assignment of material properties, and mesh generation, ensuring thorough validation and adjustments based on convergence criteria. The final output includes thermal conductivity (calculated using $\psi$-values obtained in the thermal analysis), displacement, crack width, stress, and strain.

\begin{figure}[h]
	\centering
	\includegraphics[width=0.8\textwidth]{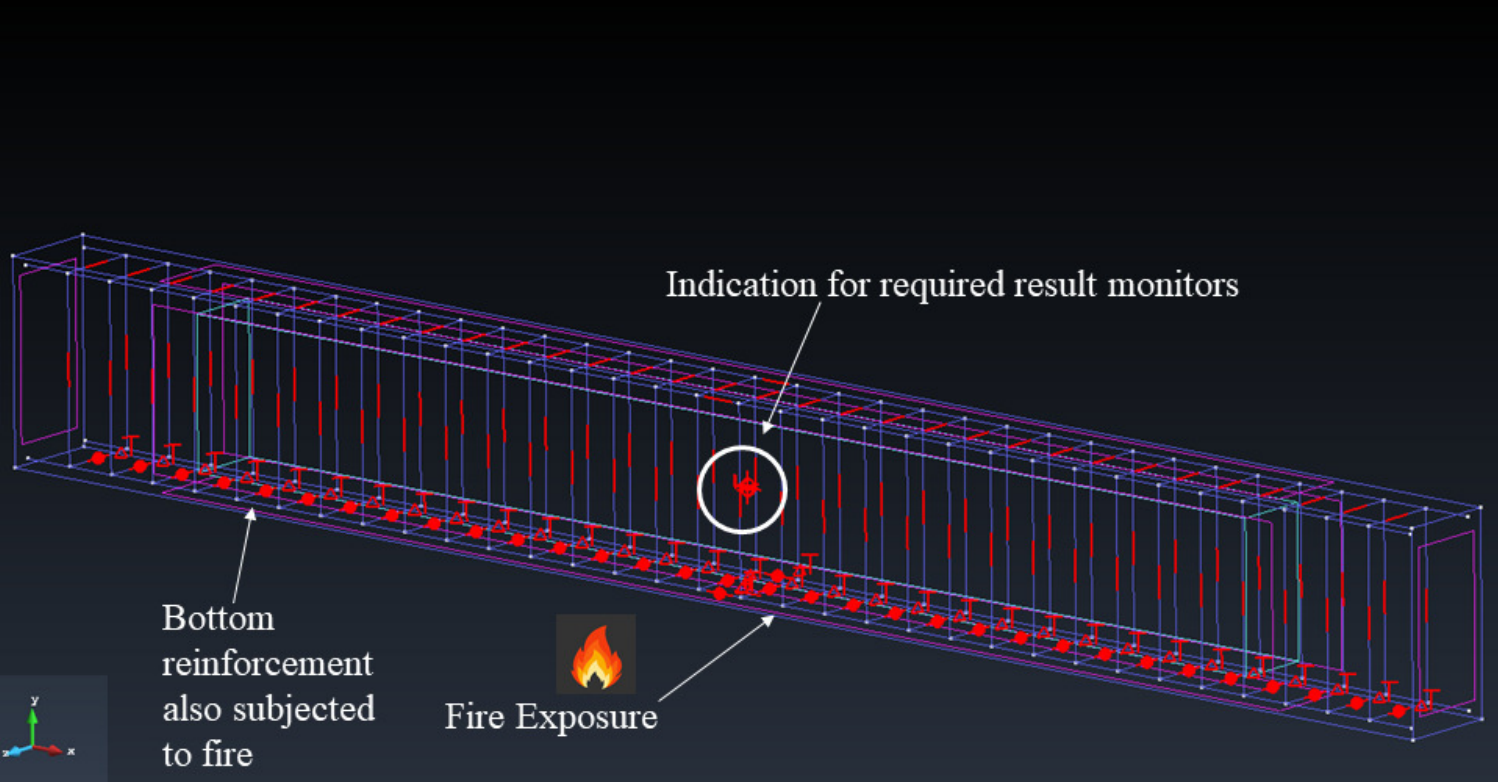}
	\caption{External loading conditions where bottom surface of beam and bottom reinforcement is exposed to fire}
	\label{figure3}
\end{figure}

\subsection{Description of external loading, material variation, and expected parameters}
In fire analysis, the main concerns are the intensity of the fire and the duration of exposure \cite{lucio2021thermal,kodur2011effect,arioz2007effects}. Fire loading is applied by selecting the surface boundary condition, with the bottom surface exposed to fire. ATENA provides this condition under the "\textit{fire boundary for surface}" setting, and for this problem, the default convection value of $50;W/cm^2$ is used. With an ambient temperature of $25^\circ$C, the maximum temperature is varied as per analysis requirements to $100^\circ$C, $250^\circ$C, $500^\circ$C, and $750^\circ$C, respectively. The ATENA model exposed to fire loading is illustrated in Fig. \ref{figure3}, showing that along with the bottom surface of the beam, the bottom reinforcement is also subjected to fire (as in the case of RCC beams).

\begin{table}[h]
	\centering
	\caption{Grade specifications from concrete mix design used in ATENA (conforming to EN 206-1)}
	\begin{tabular}{|c|c|c|c|c|c|}
		\hline
		\multirow{3}{*}{\textbf{Concrete grade}} &  & \multicolumn{2}{c|}{\textbf{Volume (m\textsuperscript{3})}} && \\
		\cline{2-6}
		& \textbf{Water/cement Ratio} & \textbf{Cement (clinker + gypsum)} & \textbf{Cement (SCMs)} & \textbf{Water} & \textbf{Aggregate} \\
		[0.5cm]	\hline
		M25 & 0.55 & 0.05435 & 0.03125 & 0.1375 & 0.7271 \\
		[0.5cm]	\hline
		M35 & 0.50 & 0.06522 & 0.0375 & 0.15 & 0.6975 \\
		[0.5cm]	\hline
		M45 & 0.50 & 0.07609 & 0.04375 & 0.175 & 0.6554 \\
		[0.5cm]	\hline
		M50 & 0.45 & 0.08696 & 0.05 & 0.18 & 0.6332 \\
		[0.5cm]	\hline
	\end{tabular}
	\label{gradespecs}
\end{table}

The material for analysis in ATENA is considered as solid concrete, designed using the conmix (concrete mix) design feature of the software. Table \ref{gradespecs} shows the specifications used in ATENA/transport in terms of volume for the concrete mix design of the grades considered. In ATENA/static, these properties, along with the Young’s modulus ($E$), tensile strength ($F_t$), compressive strength ($F_c$), and the onset of crushing ($F_{c_0}$, representing the onset of nonlinear behavior in compression), are manually entered for each grade \cite{kodur2008numerical,alvarez2013twenty,ronchi2013fire}. Several convergence criteria are set and checked after each simulation to validate the analysis. Detailed explanations of these parameters are shown in Table \ref{convergencetable}.

 \begin{table}[h!]
 	\centering
 	\caption{Description for convergence parameter used in fire analysis}
 	\begin{tabular}{|>{\raggedright}m{3cm}|>{\raggedright}m{2cm}|>{\raggedright\arraybackslash}m{10cm}|}
 		 	 		\hline
 		\textbf{Name} & \textbf{Value} & \textbf{Parameter definition} \\
 		\hline
 		Displacement Error & 0.01\% & Displacement error is based on vector norms. It is a measure of how accurately the computed displacements from the FEM simulation match the exact or reference displacements. This concept provides a quantifiable way to assess the accuracy of computed displacements, guiding improvements in modelling, meshing, and solution strategies. \\
 		\hline
 		Residual Error & 0.01\% & Residual force error is a measure of the imbalance of forces at the nodes after solving for displacements. It provides an indication of how well the equilibrium conditions are satisfied in the numerical solution. Residual force error can also be evaluated using vector norms to quantify its magnitude. By quantifying the equilibrium conditions, it helps in identifying potential issues in the model and guides improvements in meshing, boundary conditions, and solver settings. \\
 		\hline
 		Absolute Residual Error & 0.01\% & The maximum error of residual forces provides a measure of the largest imbalance in the nodal forces after solving for displacements. This parameter is essential for convergence checks, model verification, and identifying areas requiring refinement or adjustment in the FEM model. \\
 		\hline
 		Energy Error & 0.0001\% & Unbalanced energy error is a measure used to evaluate the accuracy of the numerical solution, particularly in terms of energy conservation. This error quantifies the difference between the work done by external forces and the strain energy stored in the system and helps in diagnosing model issues, ensuring energy balance, and verifying the accuracy of the FEM results. \\
 		\hline
 	\end{tabular}
 	\label{convergencetable}
 \end{table}
 
\section{Test descriptors and response monitoring }

\begin{figure}[h]
	\centering
	\includegraphics[width=\textwidth]{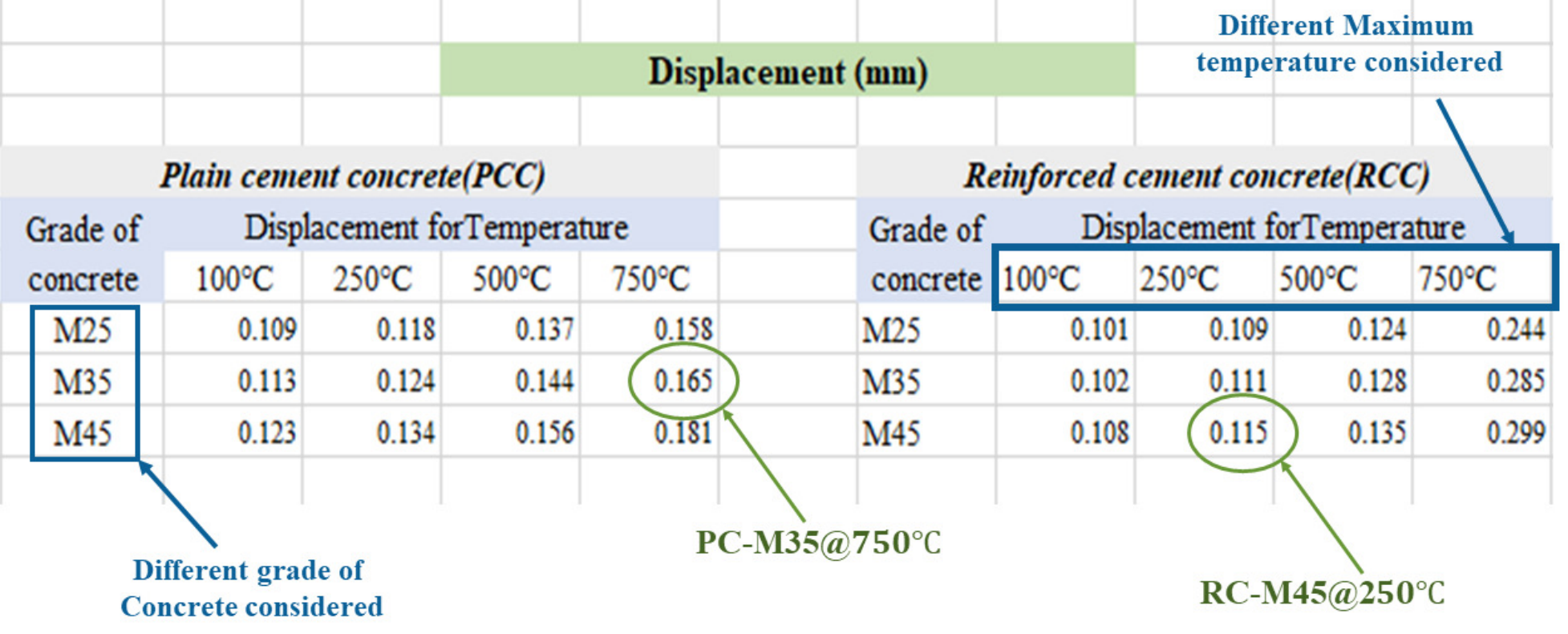}
	\caption{ Data representation of excel sheet of displacement with explanation}
	\label{figure4}
\end{figure}

The provided data is organized in an Excel file comprising five sheets, each corresponding to a specific result parameter as outlined in Table \ref{resultparam}. These parameters are recorded separately for plain cement concrete (PCC) and reinforced cement concrete (RCC).

\begin{table}[h!]
	\centering
	\caption{Result parameters}
	\begin{tabular}{|c|>{\raggedright\arraybackslash}m{8cm}|}
		\hline
		\textbf{Sheet} & \textbf{Parameter} \\
		\hline
		Sheet 1 & Thermal Conductivity (\(W/\text{m}^\circ C\)) \\
		\hline
		Sheet 2 & Displacement (\(mm\)) \\
		\hline
		Sheet 3 & Crack Width (\(mm\)) \\
		\hline
		Sheet 4 & Maximum Stress (\(MPa\)) \\
		\hline
		Sheet 5 & Maximum Strain \\
		\hline
	\end{tabular}
	\label{resultparam}
\end{table}

As illustrated in Fig. \ref{figure4}, the data is represented in Excel sheets for PCC and RCC at different maximum temperatures for the selected concrete grades. For instance, in the displacement data, the displacement in plain cement concrete for grade M35 at $750^\circ$C is $0.165$mm, marked as PC-M35@$750^\circ$C. For reinforced cement concrete, the displacement for grade M45 at $250^\circ$C is $0.115$mm, marked as RC-M45@$250^\circ$C. Similarly, other parameters across the data sheets can be analyzed and understood for both plain and reinforced concrete beams.  The data is hosted in a \textbf{GitHub} repository and can be accessed at \href{https://github.com/oscarlab-SHM/fire-in-concrete.github.io }{oscarlab-SHM}.

\begin{figure}[h]
	\centering
	\begin{minipage}{0.49\textwidth}
		\centering
		\includegraphics[width=\linewidth]{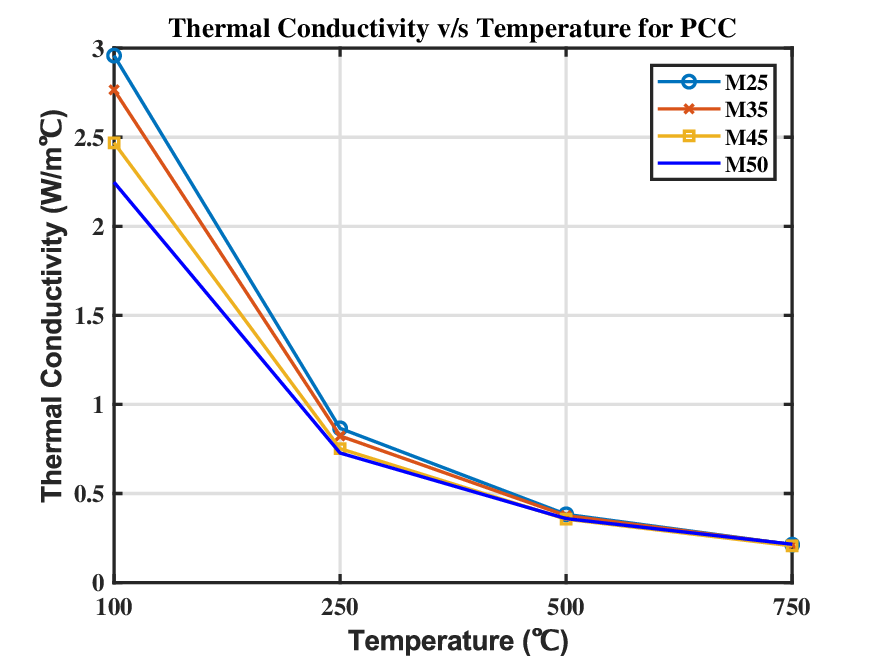} 
		\label{fig:fig1}
	\end{minipage}
	\hfill
	\begin{minipage}{0.49\textwidth}
		\centering
		\includegraphics[width=\linewidth]{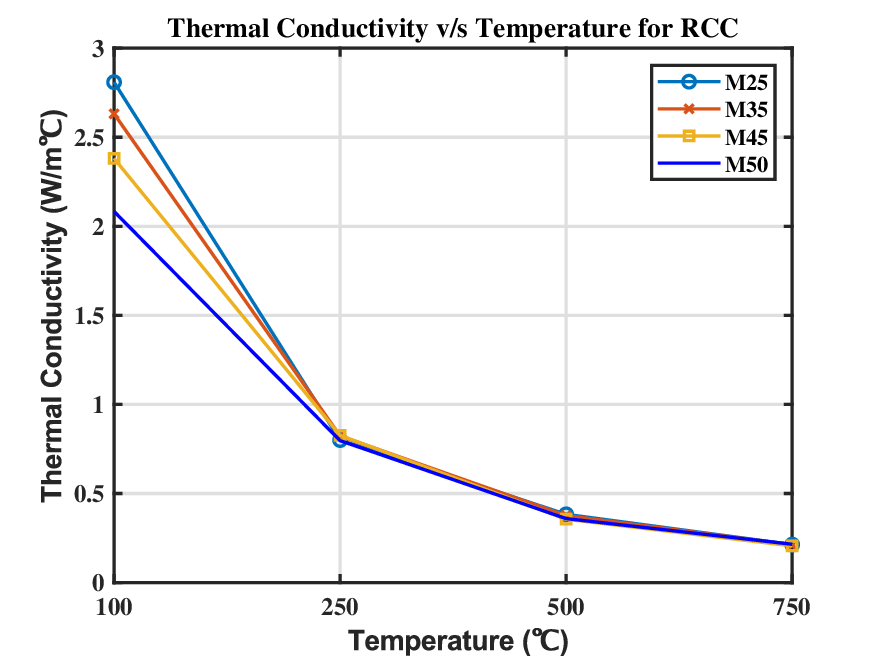} 
		\label{fig:fig2}
	\end{minipage}
	\caption{Thermal conductivity for PCC and RCC with temperature } 
	\label{Figure 5}
\end{figure}


\section{Parametric evolution of test descriptors}

The results from the fire analysis illustrate the performance of typical PCC (Plain Cement Concrete) and RCC (Reinforced Cement Concrete) beams under severe fire conditions. These findings are crucial for evaluating and advancing strategies to address challenges in fire-prone structures.  

\begin{figure}[h]
	\centering
	\includegraphics[width=0.8\textwidth]{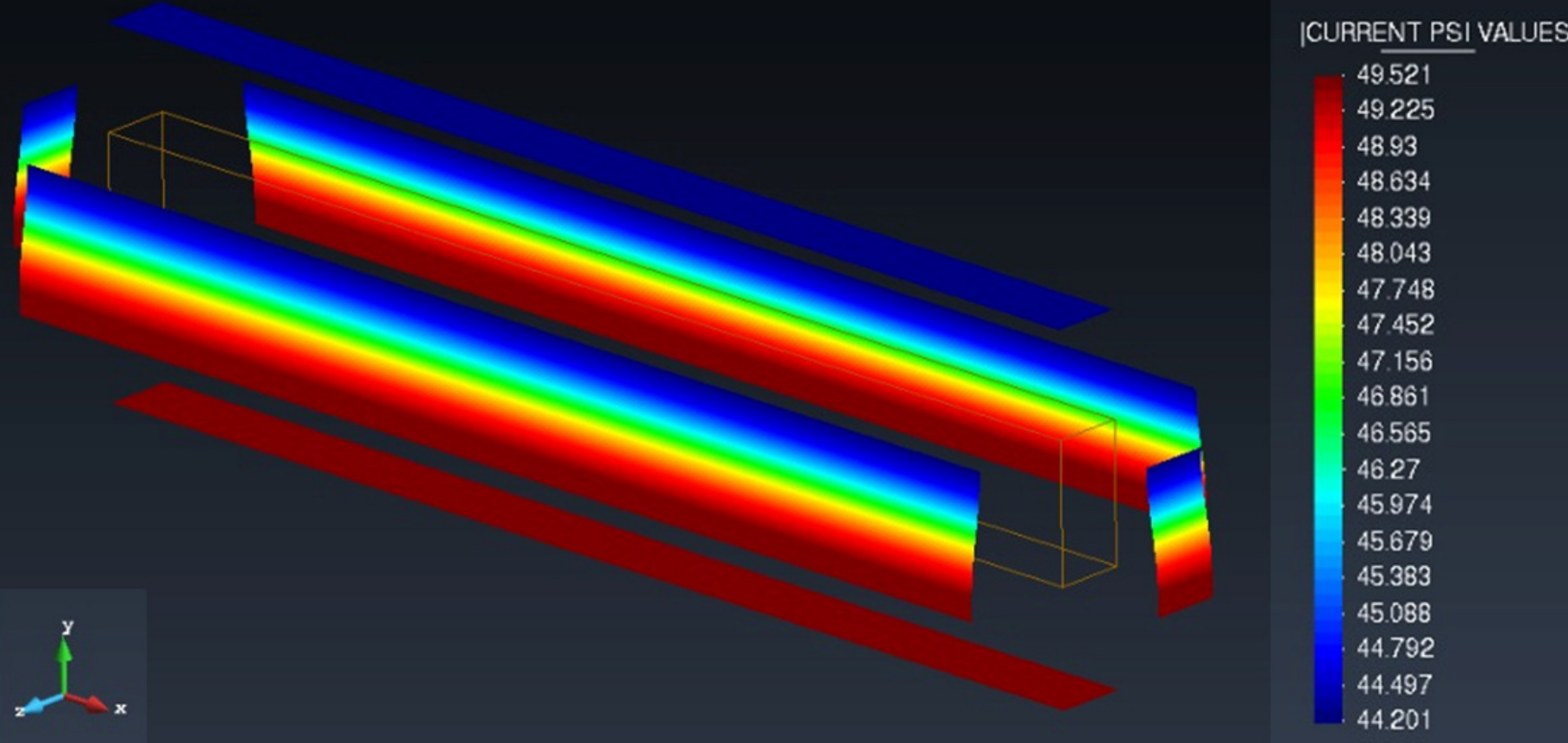}
	\caption{FEM model representation of thermal results for RC-M35 at $500^\circ$C}
	\label{figure6}
\end{figure}

\subsection{Thermal Conductivity}

Thermal conductivity is characterized by temperatures observed at various depths within the concrete and the reinforcement bars, denoted by $\psi$-values indicating heat retention in the members. Fig. \ref{Figure 5} presents these values across all concrete grades and types. As heat permeates deeper into the concrete, it affects its strength and stiffness due to increased temperatures. Higher-grade concrete exhibits lower thermal conductivity compared to lower grades, and reinforced concrete demonstrates lower thermal conductivity compared to plain concrete. The conductivity decreases with temperature due to moisture loss, particularly noticeable in higher-grade and reinforced concrete. The FEM model representation of thermal results in terms of $\psi$-values for RC-M35 at $500^\circ$C is depicted in Fig. \ref{figure6}, illustrating each surface of the beam.

\begin{figure}[h]
	\centering
	\begin{minipage}{0.49\textwidth}
		\centering
		\includegraphics[width=\linewidth]{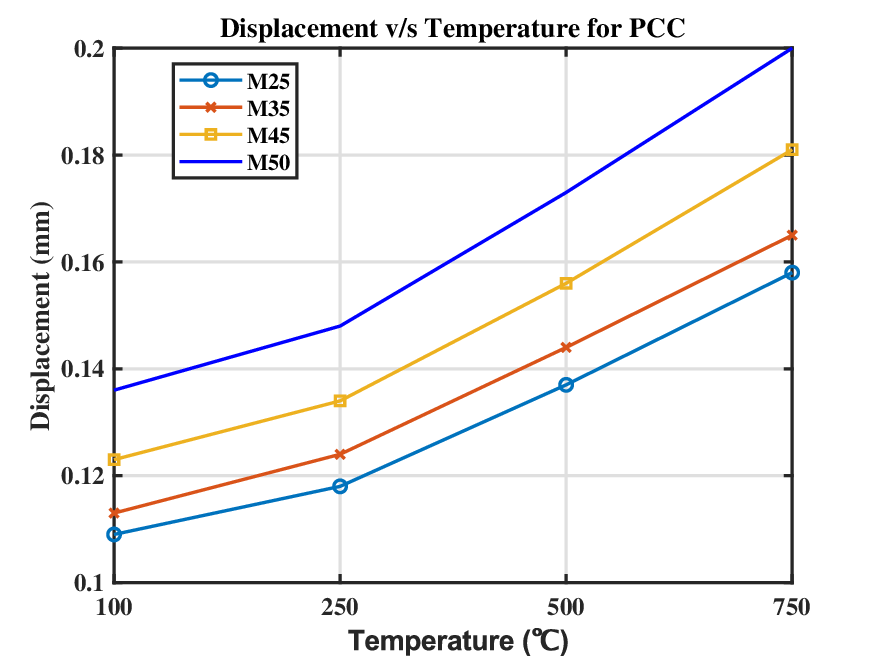} 
		\label{fig:fig1}
	\end{minipage}
	\hfill
	\begin{minipage}{0.49\textwidth}
		\centering
		\includegraphics[width=\linewidth]{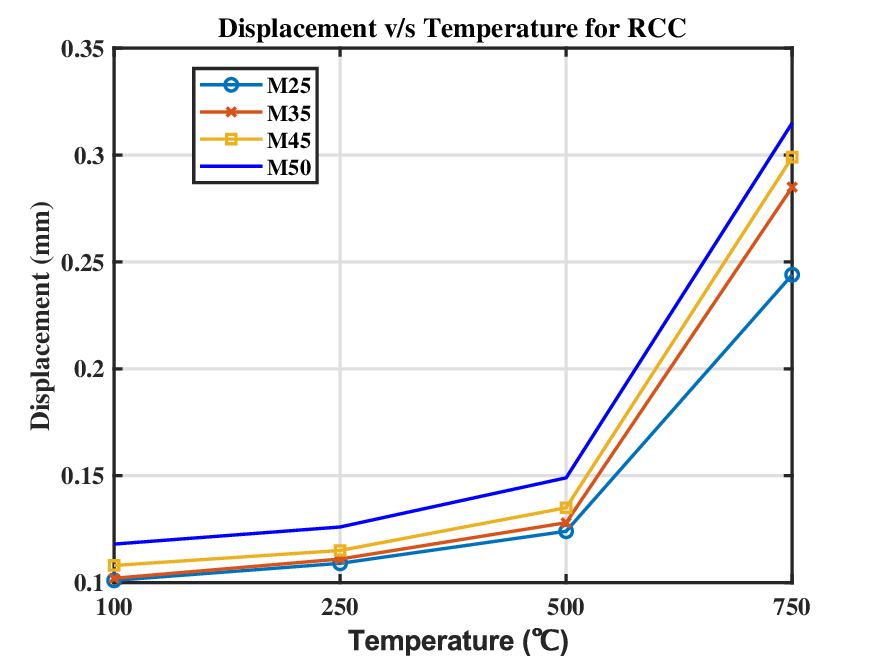} 
		\label{fig:fig2}
	\end{minipage}
	\caption{Variation in Displacement with temperature for PCC and RCC } 
	\label{Figure 7}
\end{figure}

\subsection{Displacement}
\begin{figure}[h]
	\centering
	\includegraphics[width=0.75\textwidth]{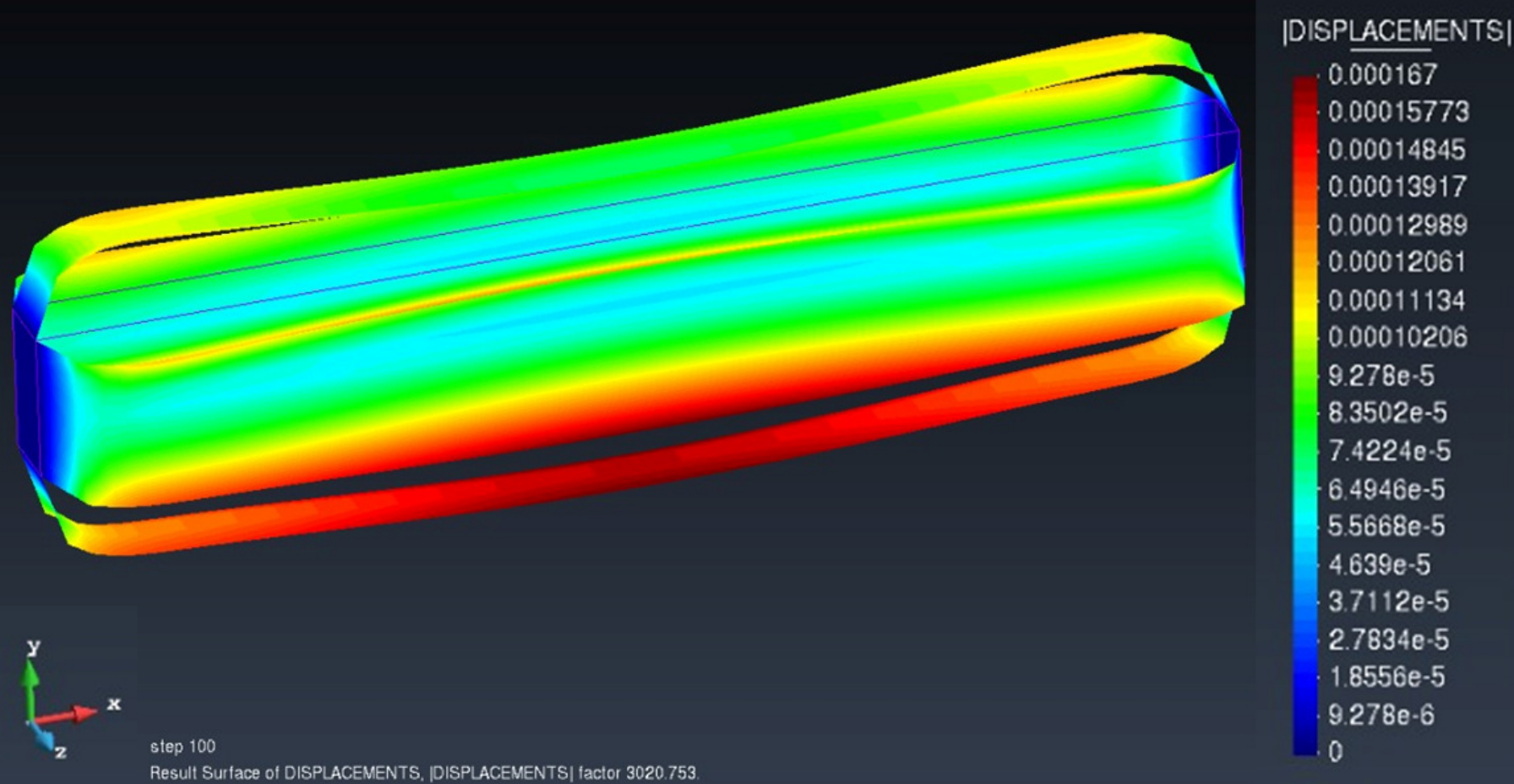}
	\caption{FEM model representation of displacement for PC-M45 at $750^\circ$C}
	\label{figure8}
\end{figure}

Higher-grade concrete typically exhibits greater displacement compared to lower grades, while RCC generally displays lower displacement than PCC due to the added strength from reinforcement. Post $500^\circ$C, RCC experiences a notable increase in displacement due to steel reinforcement losing strength. Fig. \ref{Figure 7} presents the displacement data for both concrete types across different temperatures, highlighting that beam displacement escalates with rising temperatures as material strength and stiffness degrade. The maximum deflection occurs at the mid-span of the beam in a downward direction under applied temperatures. As temperatures increase gradually, concrete initially expands at the exposed surface and begins to develop cracks between $100^\circ$C and $250^\circ$C. Fig. \ref{figure8} illustrates the displacement pattern of PC-M45 at $750^\circ$C, offering a clear visualization of displacement behavior across each face of the beam.

\begin{figure}[h]
	\centering
	\begin{minipage}{0.49\textwidth}
		\centering
		\includegraphics[width=\linewidth]{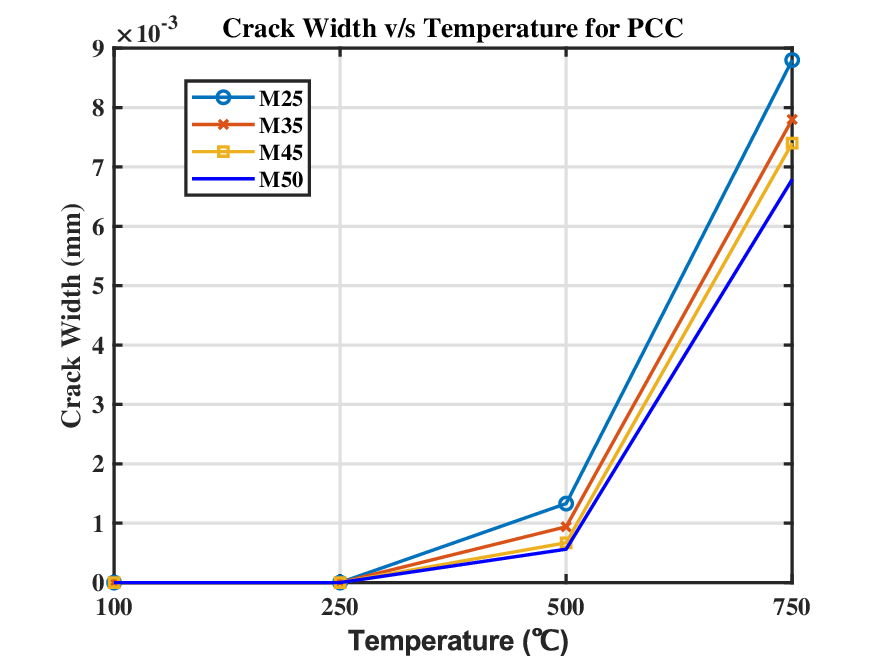} 
		\label{fig:fig1}
	\end{minipage}
	\hfill
	\begin{minipage}{0.49\textwidth}
		\centering
		\includegraphics[width=\linewidth]{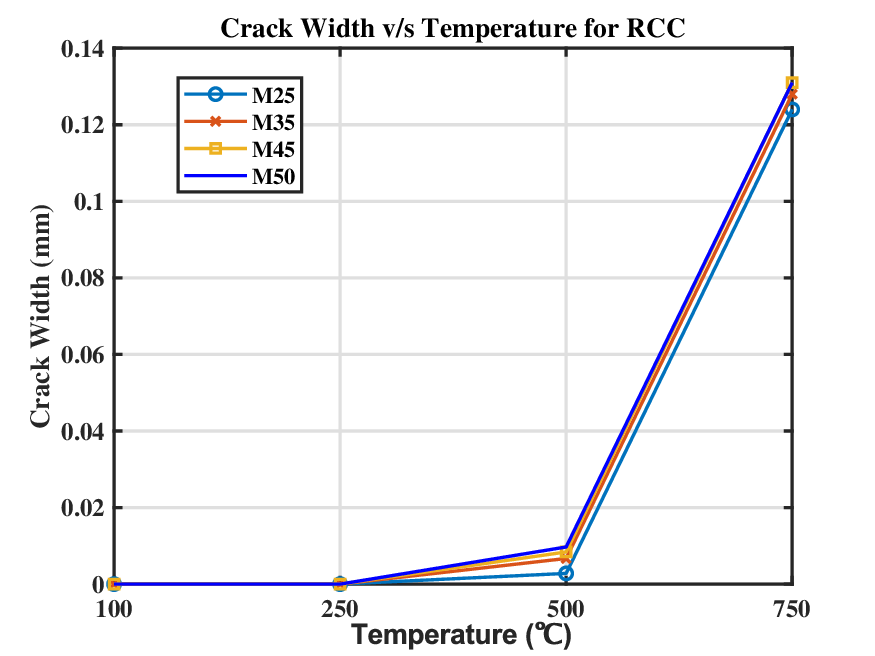} 
		\label{fig:fig2}
	\end{minipage}
	\caption{Crack width for both concrete type at different temperatures } 
	\label{Figure 9}
\end{figure}


\subsection{Crack width}
\begin{figure}[h]
	\centering
	\includegraphics[scale=0.35]{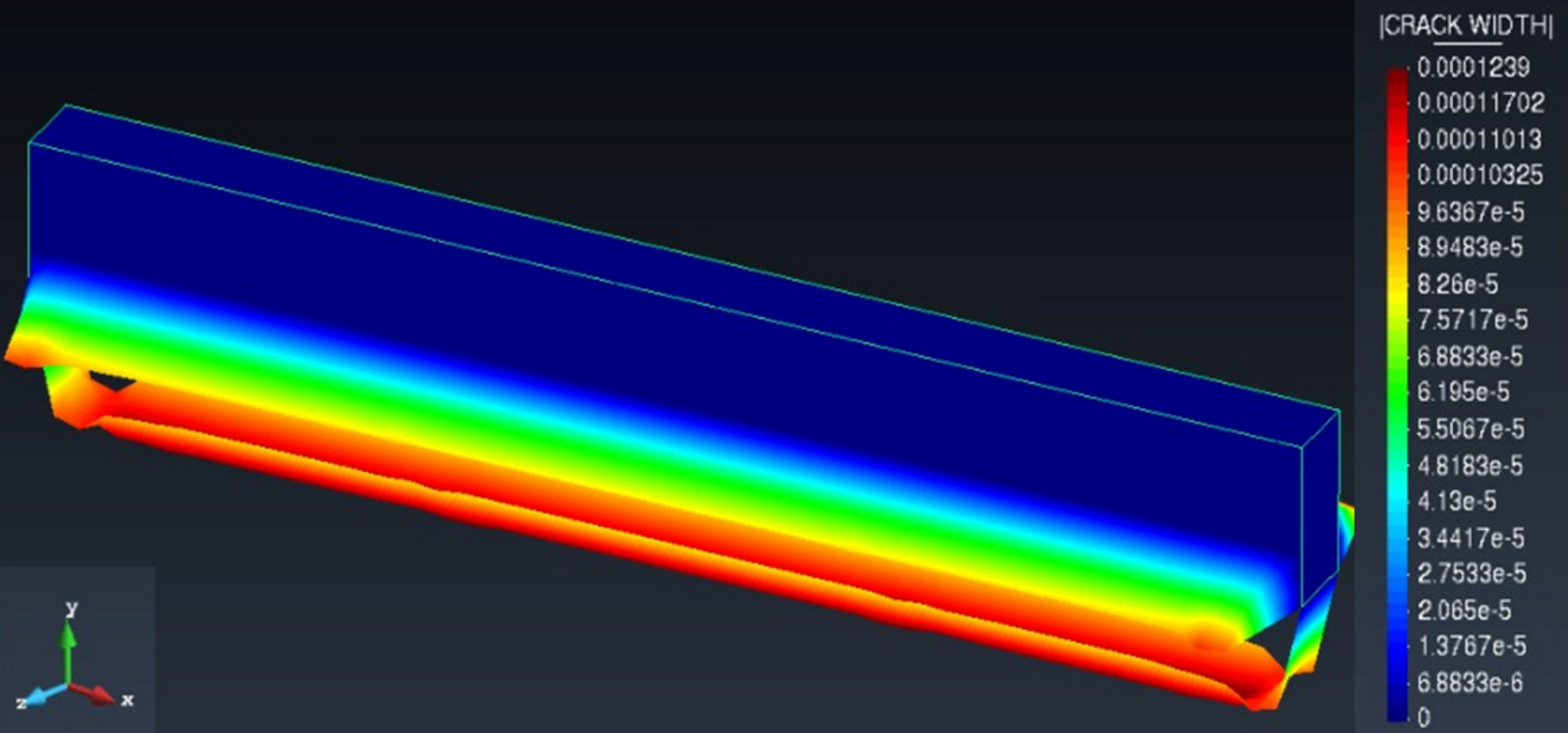}
	\caption{FEM model representation of crack width for RC-M25 at $750^\circ$C}
	\label{figure10}
\end{figure}

Crack width remains negligible in both types of beams up to temperatures of $250^\circ$C, as depicted in Fig. \ref{Figure 9}. Beyond this temperature threshold, cracks begin to gradually develop and their increase becomes more pronounced after reaching $500^\circ$C. The maximum crack width is notably influenced by the concrete grade, with higher-grade concrete exhibiting wider cracks due to its higher density. Spalling, where pieces of concrete break off from the surface, typically occurs when temperatures exceed $350^\circ$C \cite{kodur2008numerical}. The FEM model representation of crack width for RC-M25 at $750^\circ$C is illustrated in Fig. \ref{figure10}. 
 
\begin{figure}[h]
	\centering
	\begin{minipage}{0.49\textwidth}
		\centering
		\includegraphics[width=\linewidth]{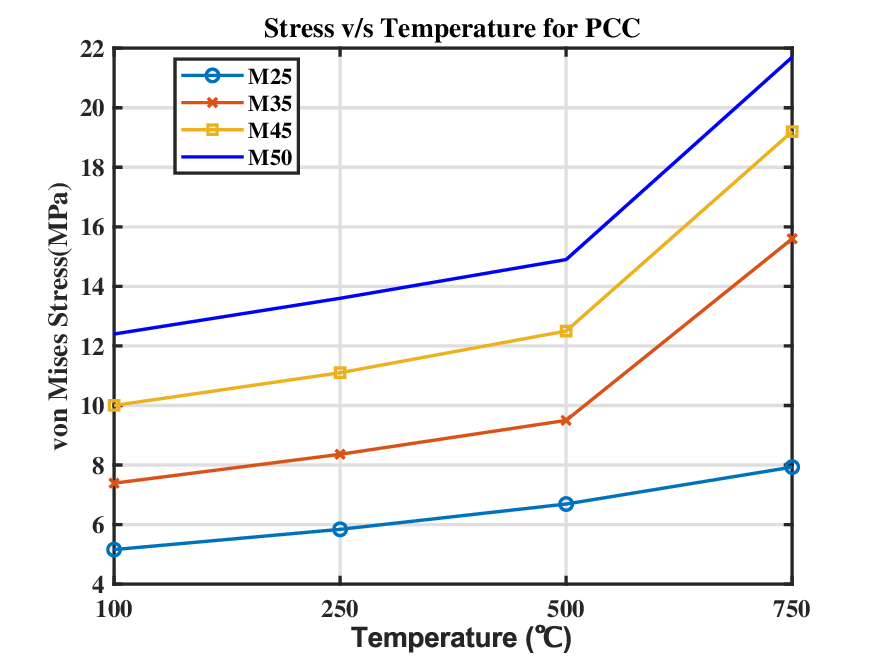} 
		\label{fig:fig1}
	\end{minipage}
	\hfill
	\begin{minipage}{0.49\textwidth}
		\centering
		\includegraphics[width=\linewidth]{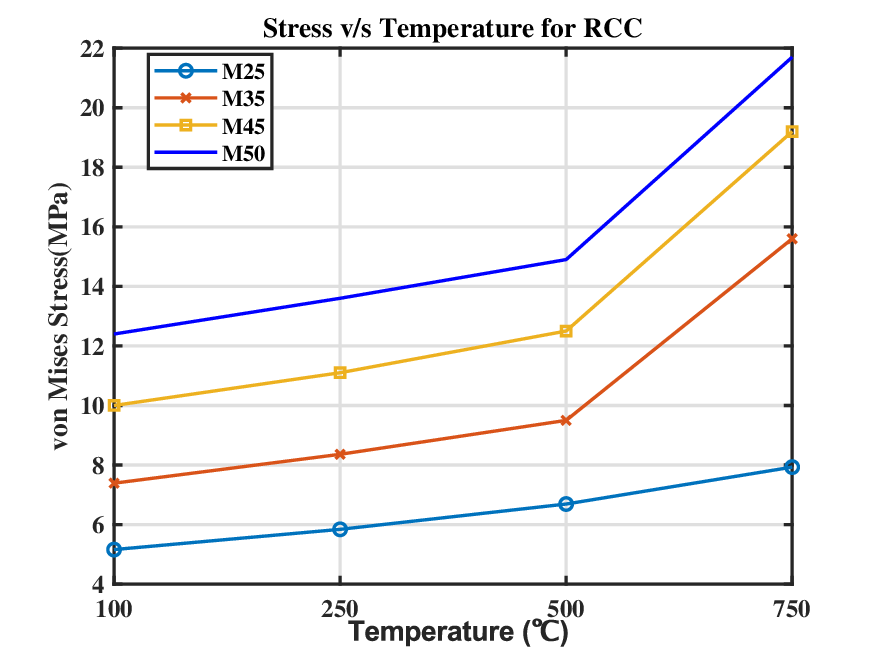} 
		\label{fig:fig2}
	\end{minipage}
	\caption{von Mises stress comparison of PCC and RCC for different grades with temperature} 
	\label{Figure 11}
\end{figure}



\subsection{Maximum Stress}
At elevated temperatures, concrete undergoes severe chemical and physical damage, resulting in expansion and spalling that can compromise its structural integrity. These physical alterations induce significant stresses within the concrete. Fig. \ref{Figure 11} demonstrates that maximum stress values increase as temperatures rise, affecting both plain and reinforced concrete similarly. Moreover, reinforced concrete beams exhibit lower stress levels compared to plain concrete beams of the same grade. This difference arises because the reinforcing steel enhances strength and resilience against thermal stresses\cite{sharma24fire}. 
 \begin{figure}[h]
 	\centering
 	\begin{minipage}{0.49\textwidth}
 		\centering
 		\includegraphics[width=\linewidth]{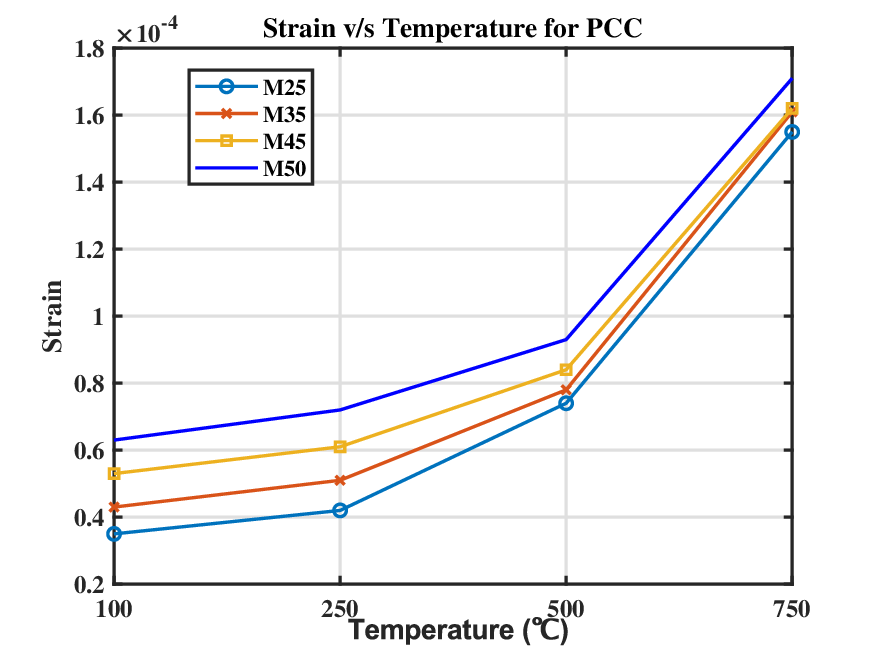} 
 		\label{fig:fig1}
 	\end{minipage}
 	\hfill
 	\begin{minipage}{0.49\textwidth}
 		\centering
 		\includegraphics[width=\linewidth]{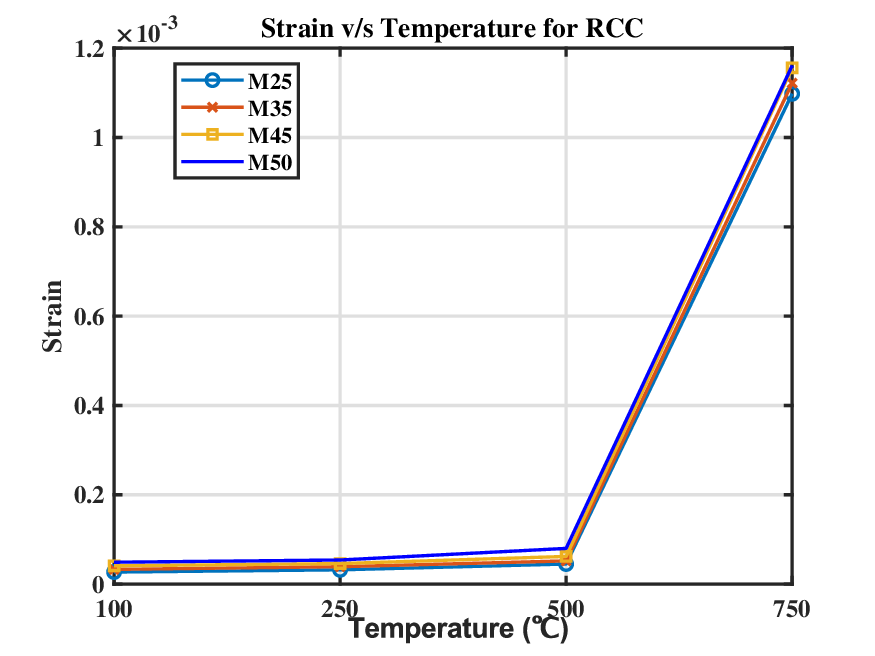} 
 		\label{fig:fig2}
 	\end{minipage}
 	\caption{Strain variation with temperature for PCC and RCC} 
 	\label{Figure 12}
 \end{figure}
%

\begin{figure}[h!]
	\centering
	\includegraphics[width=0.85\textwidth]{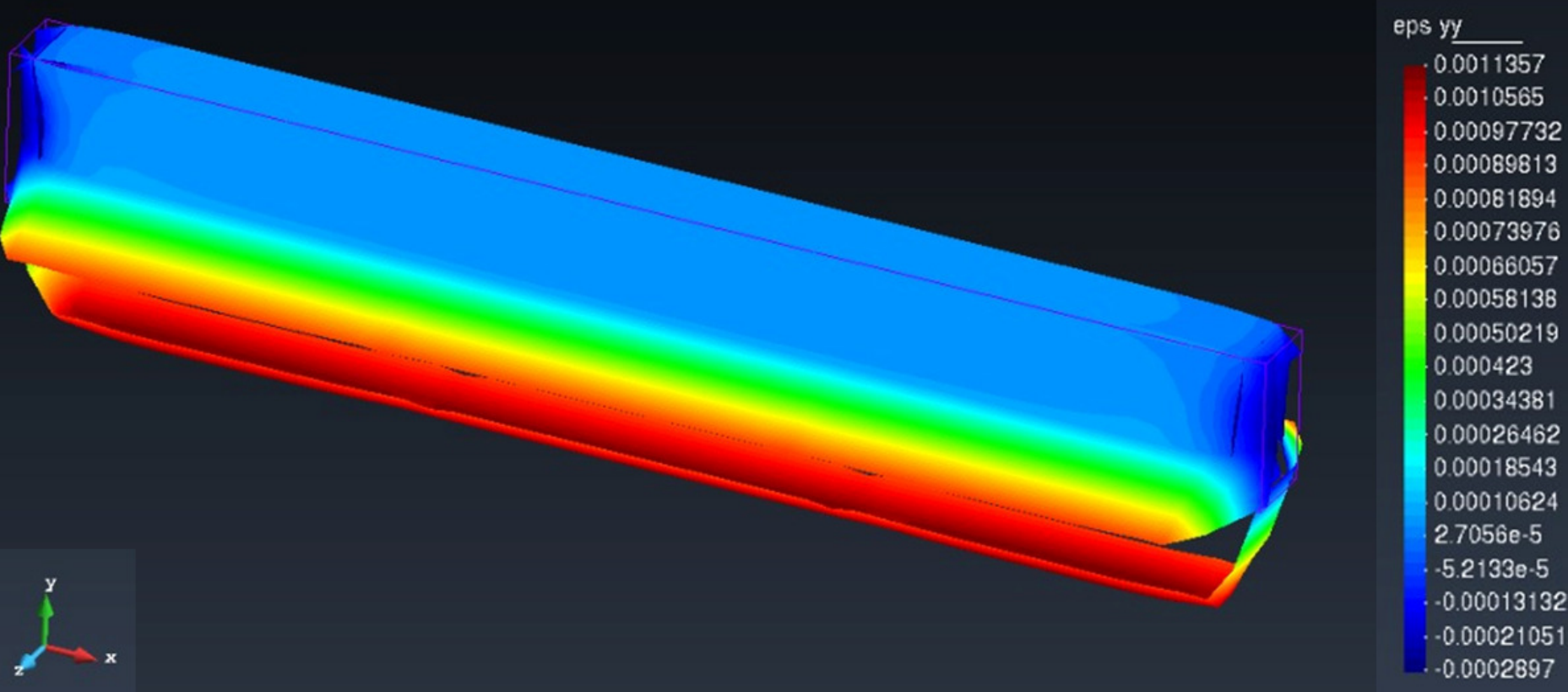}
	\caption{FEM model representation of strain for RC-M45 at $750^\circ$C }
	\label{figure14}
\end{figure}

\subsection{Maximum Strain}
In beams, strain concentration typically occurs at the mid-span initially, before any visible cracks appear. Once cracks develop, the maximum strain concentration shifts towards the spalling zone. As temperature increases, strain levels escalate for both plain and reinforced concrete. Fig. \ref{Figure 12} illustrates that plain concrete beams generally exhibit higher strain values compared to reinforced concrete beams. In RCC structures, strain remains minimal up to $250^\circ$C. However, as temperatures exceed this threshold, strain gradually increases. This distinct strain behavior is attributed to the reinforcement in RCC, which effectively resists strain until the concrete reaches higher temperatures where even the steel reinforcement starts to lose strength. The FEM model representation of strain for RC-M45 at $750^\circ$C is depicted in Fig. \ref{figure14}.
 
\section{Conclusions}
 Based on the comprehensive analysis of thermal, mechanical, and structural responses of both plain and reinforced concrete beams under severe fire conditions, several key conclusions can be drawn from this study:
 \begin{enumerate}

	\item \textbf{Thermal Conductivity and Displacement}: Higher-grade concrete demonstrates lower thermal conductivity and higher displacement compared to lower grades, while reinforced concrete (RCC) exhibits lower displacement due to enhanced reinforcement. As temperatures increase, both thermal conductivity and displacement escalate, influencing structural performance significantly.

	\item \textbf{Crack Formation and Width}: Crack initiation in concrete begins after $250^\circ$C, with higher-grade concrete showing wider cracks due to its denser composition. Spalling becomes pronounced beyond $350^\circ$C, compromising structural integrity and safety.

	\item \textbf{Stress and Strain Distribution}: Maximum stress levels increase with temperature rise, affecting both plain and reinforced concrete. However, reinforced concrete beams display lower stress concentrations due to the ability of the reinforcing steel to mitigate thermal stresses. Strain remains minimal in RCC up to $250^\circ$C, highlighting its resilience compared to plain concrete under fire conditions.

	\item \textbf{Structural Integrity and Performance}: Concrete undergoes significant chemical and physical changes at elevated temperatures, leading to expansion, spalling, and increased stresses. The presence of reinforcement in RCC delays strain accumulation and enhances overall structural resilience up to higher temperatures.

	\item \textbf{Implications for Fire-Prone Structures}: The critical importance of considering material properties and reinforcement strategies in designing fire-resistant structures is discussed in detail. Towards this, vibration-based infrastructure monitoring strategies can be designed -- keeping fire safety as the central pivot \cite{bhowmik2024advancements,bhowmik2021improved,pathak2023cnn}. Design modifications should focus on optimizing concrete composition and reinforcing techniques to mitigate the adverse effects of fire-induced thermal and mechanical stresses.

\end{enumerate}

 \section*{Acknowledgment}
 Basuraj Bhowmik gratefully acknowledges the financial support received in the form of Seed Grant from the Indian Institute of Technology (BHU), Varanasi, to carry out this research work. 
 
 \section*{Data Availability}
 Some of the data used in the paper is available at: \href{https://github.com/oscarlab-SHM/fire-in-concrete.github.io }{oscarlab-SHM}
%


\newcommand{\noopsort}[1]{} \newcommand{\printfirst}[2]{#1}
  \newcommand{\singleletter}[1]{#1} \newcommand{\switchargs}[2]{#2#1}
\begin{thebibliography}{1}
\bibitem{kodur2008numerical}
VKR Kodur and M~Dwaikat.
\newblock A numerical model for predicting the fire resistance of reinforced
concrete beams.
\newblock {\em Cement and Concrete Composites}, 30(5):431--443, 2008.

\bibitem{alvarez2013twenty}
A~Alvarez, BJ~Meacham, NA~Dembsey, and JR~Thomas.
\newblock Twenty years of performance-based fire protection design: challenges
faced and a look ahead.
\newblock {\em Journal of Fire Protection Engineering}, 23(4):249--276, 2013.

\bibitem{steen2021learning}
Anne Steen-Hansen, Karolina Storesund, and Christian Sesseng.
\newblock Learning from fire investigations and research--a norwegian
perspective on moving from a reactive to a proactive fire safety management.
\newblock {\em Fire safety journal}, 120:103047, 2021.

\bibitem{kobes2010building}
Margrethe Kobes, Ira Helsloot, Bauke De~Vries, and Jos~G Post.
\newblock Building safety and human behaviour in fire: A literature review.
\newblock {\em Fire Safety Journal}, 45(1):1--11, 2010.

\bibitem{thevega2022fire}
T~Thevega, JASC Jayasinghe, Dilan Robert, CS~Bandara, Everson Kandare, and
Sujeeva Setunge.
\newblock Fire compliance of construction materials for building claddings: A
critical review.
\newblock {\em Construction and Building Materials}, 361:129582, 2022.

\bibitem{eurocode2002actions}
EUROCODE 1.
\newblock Actions on structures--part 1--2: general actions--actions on
structures exposed to fire, 2002.

\bibitem{kodur2010high}
Venkatesh Kodur, Mahmud Dwaikat, and Rustin Fike.
\newblock High-temperature properties of steel for fire resistance modeling of
structures.
\newblock {\em Journal of Materials in Civil Engineering}, 22(5):423--434,
2010.

\bibitem{mroz2016material}
Katarzyna Mr{\'o}z, Izabela Hager, and Kinga Korniejenko.
\newblock Material solutions for passive fire protection of buildings and
structures and their performances testing.
\newblock {\em Procedia Engineering}, 151:284--291, 2016.

\bibitem{malik2021thermal}
Manisha Malik, SK~Bhattacharyya, and Sudhirkumar~V Barai.
\newblock Thermal and mechanical properties of concrete and its constituents at
elevated temperatures: A review.
\newblock {\em Construction and Building Materials}, 270:121398, 2021.

\bibitem{lucio2021thermal}
T~Lucio-Martin, M~Roig-Flores, M~Izquierdo, and M~Cruz Alonso.
\newblock Thermal conductivity of concrete at high temperatures for thermal
energy storage applications: Experimental analysis.
\newblock {\em Solar Energy}, 214:430--442, 2021.

\bibitem{kodur2011effect}
Venkatesh Kodur and Wasim Khaliq.
\newblock Effect of temperature on thermal properties of different types of
high-strength concrete.
\newblock {\em Journal of materials in civil engineering}, 23(6):793--801,
2011.

\bibitem{arioz2007effects}
Omer Arioz.
\newblock Effects of elevated temperatures on properties of concrete.
\newblock {\em Fire safety journal}, 42(8):516--522, 2007.

\bibitem{ronchi2013fire}
Enrico Ronchi and Daniel Nilsson.
\newblock Fire evacuation in high-rise buildings: a review of human behaviour
and modelling research.
\newblock {\em Fire science reviews}, 2:1--21, 2013.

\bibitem{bhowmik2024advancements}
Basuraj Bhowmik.
\newblock Advancements in online modal identification: A recursive simultaneous
diagonalization comprehensive framework for real-time applications.
\newblock {\em Engineering Structures}, 305:117770, 2024.

\bibitem{bhowmik2021improved}
Basuraj Bhowmik.
\newblock Improved single-sensor-based modal identification using singular
spectrum analysis.
\newblock In {\em International Conference on Civil Engineering Trends and
	Challenges for Sustainability}, pages 875--890. Springer, 2021.

\bibitem{pathak2023cnn}
Ishan Pathak, Ishan Jha, Aditya Sadana, and Basuraj Bhowmik.
\newblock CNN-based structural damage detection using time-series sensor data.
\newblock {\em arXiv preprint arXiv:2311.04252}, 2023.

\bibitem{sharma24fire}
Anshu Sharma and Basuraj Bhowmik.
\newblock A position paper for explainable investigations on thermal and structural behavior of concrete under extreme fire loading
\newblock {\em 10.2139/ssrn.4901385}, 2024.
\end{thebibliography}

\end{document}